\documentclass[twocolumn,showpacs,preprintnumbers,amsmath,amssymb,superscriptaddress,10pt,aps,prb]{revtex4-2}
\usepackage{epsfig,amsopn}
\usepackage{graphicx}
\usepackage{epstopdf}
\usepackage{amsmath,amssymb}
\usepackage{amsthm}
\usepackage{float}
\usepackage{enumerate}
\usepackage{xcolor}
\usepackage[normalem]{ulem}
\usepackage[colorlinks=true,linktoc=page,linkcolor=magenta,citecolor=magenta]{hyperref}

\begin{document}
\title{Frustrated Bose ladder with extended range density-density interaction}

\author{Sourav Biswas}
\affiliation{DIPC - Donostia International Physics Center, Paseo Manuel de Lardiz{\'a}bal 4, 20018 San Sebasti{\'a}n, Spain}
\author{E. Rico}
\affiliation{DIPC - Donostia International Physics Center, Paseo Manuel de Lardiz{\'a}bal 4, 20018 San Sebasti{\'a}n, Spain}
\affiliation{EHU Quantum Center and Department of Physical Chemistry, University of the Basque Country UPV/EHU, P.O. Box 644, 48080 Bilbao, Spain}
\affiliation{Theoretical Physics Department, European Organization for Nuclear Research (CERN),1211 Geneva 23, Switzerland}
\affiliation{IKERBASQUE, Basque Foundation for Science, Plaza Euskadi 5, 48009 Bilbao, Spain}
\author{Tobias Grass}
\affiliation{DIPC - Donostia International Physics Center, Paseo Manuel de Lardiz{\'a}bal 4, 20018 San Sebasti{\'a}n, Spain}
\affiliation{IKERBASQUE, Basque Foundation for Science, Plaza Euskadi 5, 48009 Bilbao, Spain}

%\ead{tobias.grass@dipc.org}

\begin{abstract}
When hard-core bosons on a two-leg ladder get frustrated by ring exchange interactions, the elusive d-wave Bose liquid (DBL) can be stabilized, a bosonic analog of a correlated metal.
Here, we analyze the effect of extended Hubbard interactions on the DBL phase. Strikingly, these interactions are found to act in favor of the exotic Bose liquid. This observation is of immediate relevance for physical systems in which non-local exchange processes occur as a consequence of extended-range density-density interactions. Our observation also helps to achieve DBL physics in a synthetic-dimension ladder, where on-site interactions translate into non-local interactions along a synthetic rung. In this context, we also consider the extreme limit, in which the local hardcore constraint is elevated to an effective rung blockade. In addition to the enhancement of DBL physics due to extended-range density-density interactions, we also find signatures of an interesting intermediate phase between the superfluid and the DBL regime. This phase, labeled as the density modulated s-wave paired (DMSP) phase, combines features of density wave and s-wave pairing. Our results offer new insights into the physics of frustrated bosons by highlighting the influence of density-density interaction and rung-blockade. 

\end{abstract}

\maketitle

\section{Introduction}\label{Intro}

\begin{figure}[t]
    \centering
    \includegraphics[width=0.475\textwidth]{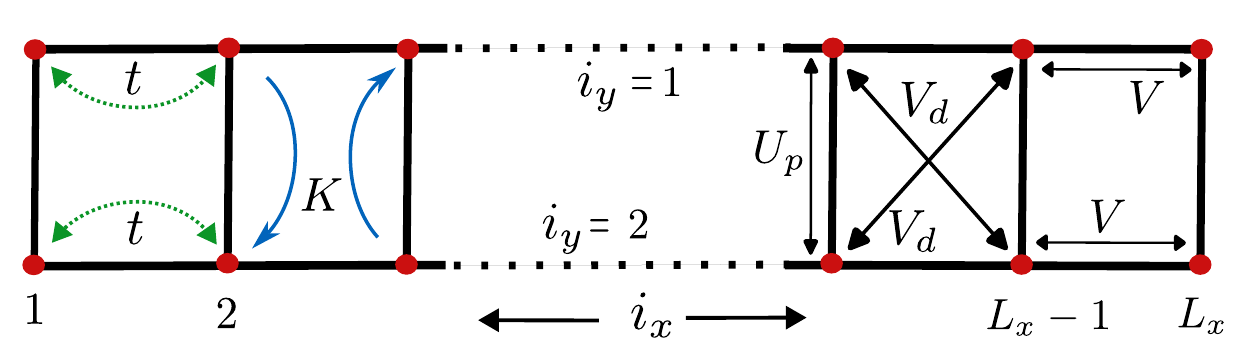}
    \caption{We base our work on the two-leg ladder model shown above. Apart from horizontal hopping $t$ and ring exchange $K$, nearest neighbor density-density interactions are also taken into account. These are shown as: $V$ for the horizontal in-chain interaction, $V_d$ represents the diagonal ones, and $U_p$ models the vertical interaction strength. } 
    \label{fig:Setup}
\end{figure}

The search for quantum phases of matter beyond the Landau phase transition paradigm has never stopped intriguing modern physicists \cite{sachdev2023quantum, fradkin2013field, coleman2015}. The non-Fermi liquids (NFLs) \cite{lee2018, Schofield, Senthil_2008, Debanjan_2018} lie at the center of this quest. In particular, several spin liquid states \cite{Balents2010, Zhou_2017} studied over the years have enriched our understanding of quantum matter. The gapless spin liquids among different types of such phases have been fascinating due to their critical nature, showing power-law decay of correlations and gapless excitations. This critical behavior can intensify the characterization of these states \cite{Senthil_2008, Hermele_2007}, however, in systems like spin-Bose metal (SBM) \cite{BlockFisher_2011, Sheng_2009} the correlations oscillate at a wave vector limited to a discrete set, corresponding to the singularities of the momentum distribution function. This can be related to the fact that this class of matter possesses singular surfaces in the momentum space, and the low energy theory is not described by a weakly interacting quasi-particle picture. In two dimensions \cite{Motrunich_2007}, it has been shown that the D wave correlated critical Bose liquids can show such features, implying the metallic behavior of bosons, which is different from the superfluid phase. It has the characteristics of the so-called Bose metal (BM) \cite{ Paramekanti_2002, Phillips_2003, Motrunich_2007, Sheng_2008, Mishmash_2011, Han_2022}.

The BM has been known theoretically for over two decades before recent experimental validation \cite{Yang_2019, Hegg_2021}. In two dimensions, the simplest notion of many-body phases of interacting bosons suggests that they exist either in a superfluid state or in a Mott phase \cite{Fisher_1989, Greiner_2002}. The advent of BM breaks down this metal-insulator binary, as the bosons form a stable uncondensed phase. This can not be explained by spontaneous symmetry-breaking of local order parameters and does not conform to any quasi-particle description. Therefore, it is of utmost interest to inquire into the development of similar behavior directly from many-body analysis. One should note, however, that the many-body physics beyond one and quasi-one dimensions get strikingly difficult due to the absence of generically reliable methods. A way to circumvent this issue is to study certain limits of higher dimensions that capture the essence of higher dimensions. The ladder models offer such a possibility: In recent years, physicists have studied a plethora of possible phases in the context of ladders systems \cite{Block_2011, Mishra_2012,Mishra_2013,Mishra_2014,Dalmonte_Lad_2023,Gia_Lad_2023,Luca_Lad_2023,Yuma2024}, including also 
BM behavior in ladders \cite{Sheng_2008, Mishmash_2011}. These studies established that the presence of ring exchange interactions produces the so-called d-wave correlated Bose liquid (DBL), as essentially the analog of two-dimensional strongly correlated phenomena within quasi-one dimension.

At the same time, the experimental availability of ladder systems has taken immense profit from advances in the vast field of quantum engineering. Cold atoms in optical lattices have provided a flexible approach for realizing ladder models \cite{Atala2014,Sompet2022}. Synthetic dimension techniques have also been used to design ladder structures via the coupling between internal levels of atoms in a strictly one-dimensional geometry \cite{Stuhl2015,Mancini2015,Arguello-Luengo2024}. Similarly, the coupling of Wannier bands can produce a synthetic ladder geometry, as proposed recently in Ref.~\cite{Yuma2024}. The concept of a synthetic ladder dimension becomes particularly useful in trapped ions systems, where Paul traps typically align the ions in one dimension. There, a synthetic ladder structure can be obtained by exploiting long-range connectivity of ions~\cite{Grass_2015}, or through a mapping onto (at least three) internal degrees of freedom~\cite{biswas2024}.

Probably the most elusive ingredient for DBL physics are ring exchange interactions between bosons. However, different possible mechanisms that can give rise to such interactions have appeared: In lattices with dipolar bosons, realized with dipolar excitons~\cite{Lagoin_2022} and dipolar atoms ~\cite{Su2023}, the presence of extended-range density-density interactions may also lead to exchange-type interactions due to Wannier function overlaps. This effect may become particularly large in synthetic ladders obtained from a one-dimensional chain. In such scenarios, the unavoidable presence of extended-range density-density interactions may affect the DBL physics in a way that has not been explored yet. Another highly tunable route toward ring exchange interactions has recently been proposed in Ref.~\cite{biswas2024}, where the ladder is mapped to a chain of three-level ions, and appropriately chosen Raman coupling between the levels maps onto ring-exchange terms. Importantly, this mapping constrains the ladder to a maximum occupation of one boson per rung, which translates to an infinitely large density-density interaction along the rung.

Motivated by these different scenarios where ring-exchange interactions and extended-range interactions are simultaneously present in bosonic ladder systems, the present study addresses the effect of extended interaction in a frustrated ladder. We confine ourselves to extended density-density interaction on a plaquette as well as rung blockade interactions. 
%
%We base our work on the two-leg bosonic ladder, which is a minimal arrangement for studying the role of ring exchange. One can think of a two-band system realizable with dipolar atoms \cite{Ferlaino2022, Yuma2024, Dutta2015} to manifest such a mechanism. However, in physical systems, other relevant interactions might also be present. The density-density interactions, in particular, can not be neglected since these are expected to be higher in strength than the ring exchange. A primary objective of this work is to address this issue and look into the effect of extended interaction in a frustrated ladder.
%
%It is plausible to think of synthetic platforms \cite{Grass_2015, Monroe_2021, Ringbauer_2022} mimicking spin systems for materializing suitable exchange mechanisms leading to the DBL phase. A similar idea has been proposed recently in ion traps~\cite{biswas2024}. All of these works, along with ongoing progress made in the field of tunable quantum matter, including dipolar gases, have been inspiring to push the boundaries of understanding regarding the DBL phase in the light of different competing interactions.  
%
Specifically, the paper is arranged as follows: In Sec.~\ref{Model}, we discuss the microscopic Hamiltonian used in our study. In Sec.~\ref{Phases}, we explore different phases relevant for both the density-density interaction and without it, keeping in mind the two-band dipolar system. The quantities required to analyze different phases that emerge from the Hamiltonian model are discussed in Sec.~\ref{OBS}. Here, we review the relevant correlations used in this work. We elaborate on the properties of distinct phases present in our model (\ref{PhSF}, \ref{PhDBL}, \ref{PhDMSP}). In this context, we also report the existence of a novel intermediate phase between the superfluid and the DBL (\ref{PhDMSP}). This so-called density modulated s-wave paired (DMSP) phase combines features of density wave and s-wave pairing. In Sec.~\ref{DDI}, we look into the role of the density-density interaction and discuss how the transitions between different phases get modified. Here, we specifically focus on how asymmetry in density-density interaction can be used to stabilize the DBL further. In Sec.~\ref{StbDBL}, we look into the rung blockade regime, which is suitable for a synthetic ladder. In Sec.~\ref{Discsn} we summarize all the results and comment on the principal observations. Throughout this work, we have used the density matrix renormalization group (DMRG) \cite{DMRG_White, Scholl_MPS} technique for studying the many-body phases. DMRG is performed with bond dimensions between $5\times10^3$ and $10^4$ as per the convergence requirements, using ITensor library \cite{itensor}. The relative truncation error is kept at the order of $10^{-10}$. 

\begin{figure}[t!]
    \centering
    \includegraphics[width=0.475\textwidth]{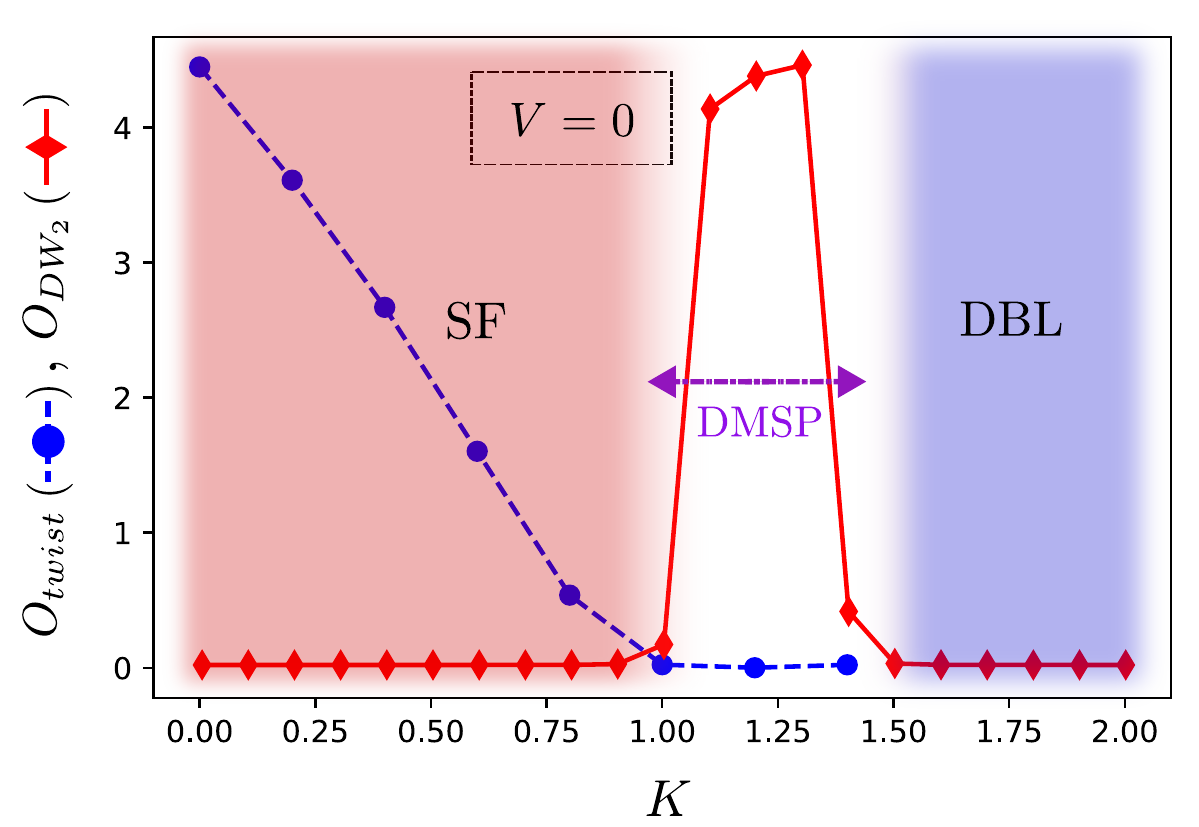}
    \caption{Different phases as a function of $K$ are shown at $V=0$. We used $L_x=36$. The rightmost ($i.e.$ DBL) shaded region has double peaks in momenta distribution. The left and middle regions cannot be distinguished based on momenta peaks. One has to look into two different order parameters $O_{twist}$ and $O_{DW_2}$. The middle part shows density wave modulation, and the left part shows finite SF stiffness.} 
    \label{fig:phase}
\end{figure}

\section{Model}\label{Model}
We consider a two-leg Bose system with $i =(i_x,i_y)$ sites, where $i_x, i_y$ represent the co-ordinates along $\hat{x}$ and $\hat{y}$ respectively. The index $i_x$ runs from $1,.., L_x$ and $i_y=1,2$. The ladder consists of $L=2L_x$ sites in total, with $N$ particles. The filling of the system is defined as $n_f = N/L$. 
%We take into account the nearest-neighbor hopping along $\hat{x}$ given by $t$, a ring-exchange term denoted by coupling $K$, and the nearest-neighbor (NN) extended interactions in all possible directions. 
Moreover, there is a ring-exchange term of strength $K$, where two particles on opposite corners of a plaquette simultaneously hop onto the other (empty) corners. In addition to this exchange-like interaction, we also consider density-density interactions. Apart from hard-core interactions on-site, there are nearest-neighbor (NN), interactions of strength $V$ along the horizontal directions and of strength $U_p$ along the vertical (rung) direction. Finally, we also consider next-nearest-neighbor (NNN) interactions of strength $V_d$ along the diagonal direction. We fix $V_d = V/(\sqrt{2})^3$. All these terms are explained in Fig.~\ref{fig:Setup}. The Hamiltonian $H$ for the system is written as
\begin{align}\label{MasHam}
H
&=H_t + H_V + H_K,
\end{align}
where the hopping is governed by the Hamiltonian 
\begin{align}
H_{t}= 
&-t \sum_{i_x, i_y} ( b^{\dagger}_{i_x+1,i_y} b_{i_x,i_y} + {\rm h.c.} ) ,
\end{align}
the NN extended interaction is governed by the Hamiltonian 
\begin{equation}
\begin{aligned}
    H_{V}
& = \sum_{i,j} \Big( V\delta_{j_x,i_x+1}\delta_{j_y,i_y} + V_d \delta_{j_x,i_x+1}(1-\delta_{j_y,i_y}) 
\\ &~~~~~ + U_p\delta_{j_x,i_x}(1-\delta_{j_y,i_y})  \Big)~  n_{i_x,i_y} n_{j_x,j_y},
\end{aligned}
\end{equation}	
and the ring-exchange mechanism is governed by the Hamiltonian  
\begin{align}
H_{K}
= K \sum_{i_x} b^{\dagger}_{i_x,2} b_{i_x,1} b^{\dagger}_{i_x+1,1} b_{i_x+1,2} .
\end{align}	

Here, the operators $b^{\dagger}_{j_x,j_y}$ create a hard-core boson at the site $j = (j_x,j_y)$, and $n_j =n_{j_x,j_y}= b^{\dagger}_{j_x,j_y} b_{j_x,j_y}$ is the local density at site $j$. 

We note that we have not included any vertical hopping in our study. It is known that such a term shifts the onset of the DBL phase further to a higher value of ring exchange strength \cite{Sheng_2008}, hence the absence of vertical hopping facilitates DBL formation. Importantly, synthetic ladder structures, including the implementation of ring exchange physics proposed in Ref.~\cite{biswas2024}, naturally avoid vertical hopping. In optical lattice systems, hopping along the rungs can be avoided through a potential mismatch between the ladders. From the theoretical point of view, an important consequence of absence of vertical hopping is the conservation of the boson occupation number in each leg separately.

\section{Phases} \label{Phases}
We begin our study of the microscopic model, described in Sec.~\ref{Model}, at the limit $V=U_p=0$, which already allows for developing a detailed understanding of the emergent phases. We fix the filling at $1/4$ and set $t=1$ for the rest of the paper unless otherwise noted. %At this level, we are motivated by a two-band model (realizable in dipolar gases, for example). In such a scenario, the $H_K$ of Eq.~\eqref{MasHam} is generated through a higher-order \cite{Halimeh2025} two-body exchange mechanism. 
%Our primary objective for the present work is to understand the role of NN and NNN interactions, which should be present if a ring-exchange term is taken into account. However, 
First, we describe the quantities of interest, based on which the characterization of different phases will be done.

\begin{figure} [t]
    \centering
    \includegraphics[width=0.45\textwidth]{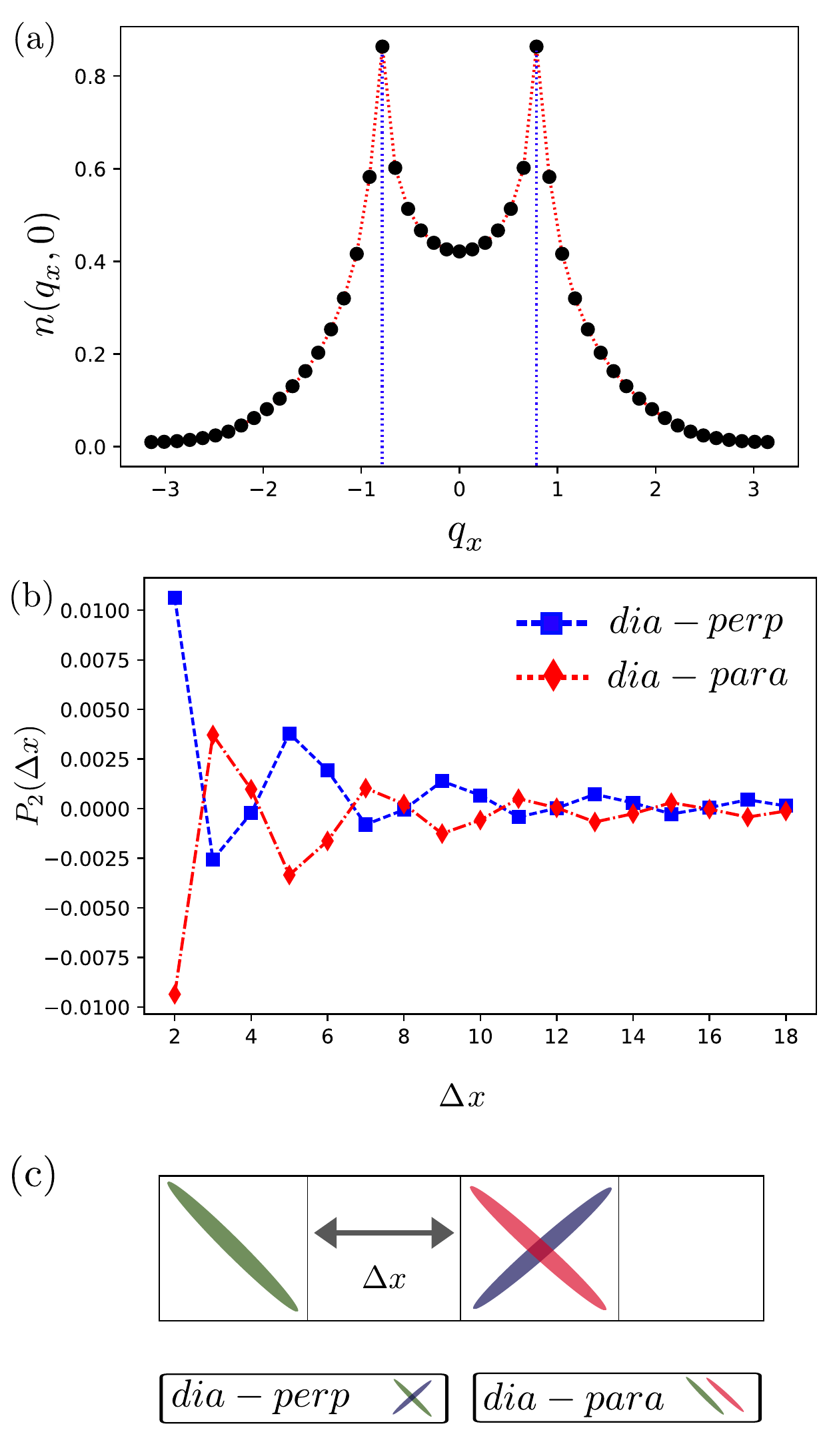}
    \caption{We show (a) momentum distribution and (b) pair correlation for the DBL phase with $L_x=48$, at $K=2.0$ and $V=0.0$. In (c), a schematic representation for the choice of diagonals is presented, which has been followed for the computation of $P_2$. In panel (a), we have indicated the position of the $q_x= \pm \pi n_f$ by dashed line, showing that momenta peaks exactly coincide with the filling. Also, the pair correlation oscillates with a period of $1/n_f=4$ sites.
    } 
    \label{fig:DBL}
\end{figure}

\subsection{Observables} \label{OBS}
The momentum distribution function becomes a powerful tool for identifying the DBL phase. This quantity is given by
\begin{align}
n(q_x,q_y) = \frac1L \sum_{i,j}  e^{-i q_x (i_x-j_x)-i q_y (i_y-j_y)} \langle b^{\dagger}_{i_x,i_y} b_{j_x,j_y} \rangle.
\end{align} 
In the absence of vertical hopping, the single-particle correlations between the legs are suppressed, and as a result $n(q_x,0) =n(q_x,\pi) $. We choose to work with $q_y=0$ only. The number and position of momenta peaks in the distribution are crucial for differentiating DBL from other phases, as we will see in the following sections of the paper. Close inspection of various functions at different parameters enables us to define $\tilde{n} (q_x,0)$ that can reliably represent the location of $n(q_x,0)$ peaks. It is given by 
\begin{align}\label{NqP}
\tilde{n} (q_x,0) = \frac{n(q_x,0)}{max(n(q_x,0))}.
\end{align}
This quantity reaches its maximum, which is exactly $1$ at the peak, and away from the peak, it varies between $0-1$.

The nature of pairing present in different phases is understood in terms of the pair correlation function
\begin{align}
P_2(\Delta x) = \langle b^{\dagger}_{1,1} b^{\dagger}_{2,2} b_{\delta_1+\Delta x, 1} b_{\delta_2+\Delta x , 2}  \rangle,
\end{align}
where $(\delta_1,\delta_2)=(2,3)$ represents two-particle correlation between parallel diagonals ($dia-para$), and $(\delta_1,\delta_2)=(3,2)$ is in between perpendicular diagonals ($dia-perp$).

Our results also suggest the presence of density wave in particular cases and the density pattern follows a two-site modulation captured by the order parameter $O_{DW_2} = \sum_{j_y} |\mathcal{O}^{j_y}_{DW_2}|$, where
\begin{align}
\mathcal{O}^{j_y}_{DW_2} = \sum_{j=1}^{L_x/2} (-1)^{j} (n_{2j-1,j_y} + n_{2j,j_y}). 
\end{align} 

The superfluid (SF) phase can be characterized by SF stiffness. We use twisted boundary conditions ($\theta$ being twist angle) on a periodic chain, following \cite{Sheng_2008}, and compute ground state energies ($E^{\theta}_{GS}$) for twist angles $\theta=0$ and $\theta=\pi$. The difference between these two energies signifies the rigidity of the phase, captured by $O_{twist} =  L_x\times (E^{0}_{GS} - E^{\pi}_{GS})$. 

We are now in a position to start looking into different phases of matter. We encounter three distinct phases as a function of $K$: ($i$) the superfluid phase (SF), ($ii$) the density wave modulated s-wave pairing phase (DMSP), and ($iii$) the d-wave correlated Bose liquid phase (DBL). This is illustrated in Fig.~\ref{fig:phase}. 

\subsection{Superfuid phase (SF)}\label{PhSF}

The SF phase shows a finite superfluid stiffness, that is, a finite value of $O_{twist}$. The momentum distribution $n(q_x,0)$ of the SF has a peak at zero momentum $q_x=0)$. The pair correlations $P_2$ decay according to a power law, establishing quasi-long-range order in 1D. There is a vast amount of literature for the SF phase in the Bose-Hubbard system, including ladders \cite{White_2000,Block_2011, Mishra_2012, Rossini_2012,Mishra_2013,Mishra_2014,Gia_Lad_2023,Luca_Lad_2023,Yuma2024}. Our observations follow the known properties of SF, and for the sake of completeness, an example of the finite stiffness is  shown in Fig.~\ref{fig:phase}. We choose to focus on the less explored phase\textit{, i.e.} the DBL and the novel DMSP.

\subsection{d-wave correlated Bose liquid phase (DBL)}\label{PhDBL}

The DBL phase is characterized by the presence of singularities at momenta $\pm \pi n_f$, clearly different from a single peak in the SF phase, see Fig.~\ref{fig:DBL}(a). The absence of a zero momenta peak signifies the absence of condensation. 
The pair correlation $P_2$ of the DBL is known to oscillate with a period $n_f^{-1}$, and pair correlations between the diagonals in parallel and perpendicular configurations are opposite to one another, see Fig.~\ref{fig:DBL}(b) and (c). This behavior establishes the ``d-wave-ness'' of this phase \cite{Mishmash_2011, Sheng_2008}. 

The DBL phase can not be explained by the spontaneous symmetry breaking of any local order parameter fields and lacks any classical order. The many-body state is dominated by quantum fluctuations, having a d-wave-like spatial dependence. The momenta peaks signal the emergence of a new scale in the system, which corresponds to new gapless modes. The system essentially goes into a phase where interacting bosons form Fermi surface-like singularities, resulting in metallic features in a system of bosons, deemed the elusive Bose metal (BM). As a result, the possibility of having a metallic Bose system becomes plausible in a model of a frustrated Bose ladder.

It is noted that $O_{twist}$, used as a marker for superfluid stiffness, can show irregular behavior in the DBL region. In this context, it is important to realize that $O_{twist}$ only makes sense in the presence of a zero momenta peak, and it should not be assigned any physical meaning if there is no peak around zero momenta, as this signals an absence of condensation.

 \begin{figure}[h!]
    \centering
    \includegraphics[width=0.45\textwidth]{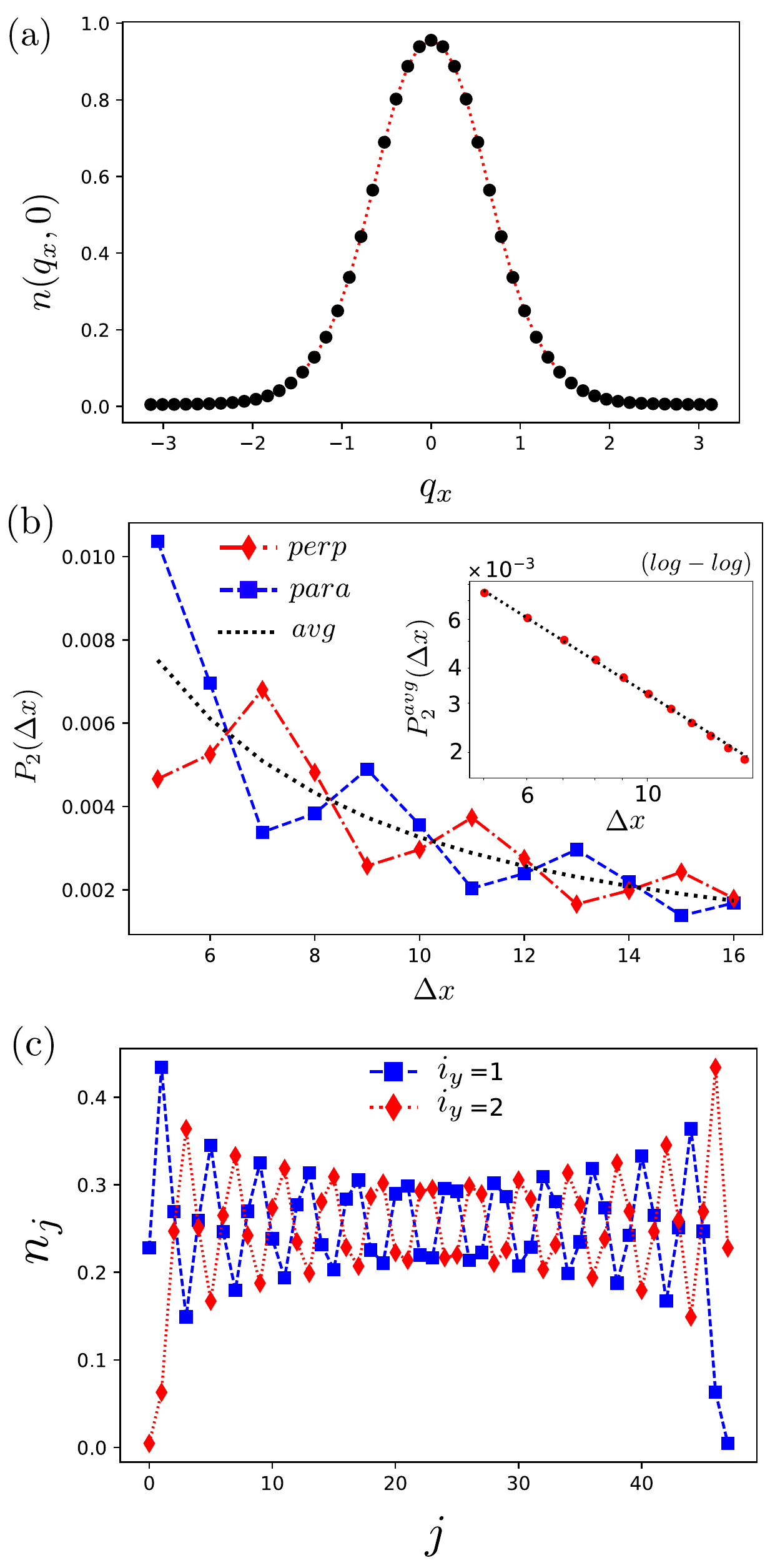}
    \caption{We show the properties of the DMSP at $K=1.2$ and $V=0.0$, with $L_x=48$. (a) The momentum distribution has a single peak. (b) The two-particle diagonal pairing shows s-wave-like behavior. The $dia-perp$ and $dia-para$ configurations (see Fig.~\ref{fig:DBL}(c)) are denoted as $perp$ and $para $, respectively. The curve marked as $avg$ shows the mean value of these two configurations, which is $P_2^{avg}$. The same is shown in the inset on a $log-log$ scale to visualize the power scaling. However, the single-particle density (c) shows density wave ordering.} 
    \label{fig:DMSP}
\end{figure} 

\subsection{Density wave modulated s-wave pairing phase (DMSP)}\label{PhDMSP}

While the DBL phase can be unambiguously identified from the momentum distribution $n(q_x,0)$, the momenta peaks are not sufficient to distinguish between the superfluid and the intermediate phase appearing in Fig.~\ref{fig:phase}. As we can see in this figure, for values of $K$ between approximately 1 and 1.5, the superfluid order is destroyed, indicated by the vanishing of $O_{twist}$. However, the splitting of the momentum distribution peak, indicative of the DBL phase, has yet not set in. Instead, the two-site density modulation order parameter, $O_{DW_2}$, which is zero in both the SF and the DBL phase, takes finite values in this intermediate regime (see red line in Fig.~\ref{fig:phase}). As a result, $O_{DW_2}$ serves as a reliable identifier for the intermediate region between SF and DBL, which we denominate as density wave modulated s-wave pairing phase (DMSP).

Other properties of the DMSP phase are shown in Fig.~\ref{fig:DMSP}: In panel (a), it is shown that from the point of view of momentum distribution, the phase resembles an SF, as it exhibits a zero-momentum peak. As shown in Fig.~\ref{fig:DMSP}(b), the pair correlation oscillates, but different from the DBL phase, there is no sign difference between perpendicular and parallel configurations. This specific nature of the two-particle pairing is known to be s-wave \cite{Sheng_2008}. The decay of the correlations indicates a quasi-long-range order, as expected in lower dimensions. This can be understood by plotting $P_2^{avg} = (P_2^{perp}+P_2^{para})/2$. The correlations for perpendicular (parallel) configuration of the diagonals, $i.e.$ $P_2^{perp}(P_2^{para})$ oscillate about $P_2^{avg}$. This quantity is shown to follow a power-law in the inset of Fig.~\ref{fig:DMSP}(b). \\ 
We have also found that the ground state is two-fold degenerate, and we use appropriate pinning potential to break the degeneracy, obtaining a density-wave modulation with an open boundary. We show the modulation of the density of the ground state in Fig.~\ref{fig:DMSP} (c). One can compute the exact same quantity for the degenerate counterpart and check that the density modulation in each of the legs is exactly opposite for the two orthogonal ground states. This reflects the presence of a $\mathbf{Z}_2$ symmetry. Hence, the degenerate ground states show density wave patterns at the single-particle level and pairing signatures at the two-particle level. This intermediate phase exists at different values of $V$, as we will see later. \\
In the past, an intermediate phase between SF and DBL, with s-wave pairing \cite{Sheng_2008}, has been reported. Our results agree with the findings therein, in terms of the pairing correlation and momenta distribution. However, the referred work did not delve into searching for density modulation or degeneracy, as per our understanding. In that regard, there is no conflict with the known properties of this intermediate phase. However, observing degenerate states with density modulation provides a new perspective on the physics of ring exchange.\\ 
We end this section by summarizing the properties of different phases in terms of correlators discussed in sec-\ref{OBS}. The Table-\ref{Tab:Ph} is referred to. We note that the quantity $sgn(P^{perp}_2)/sgn(P^{para}_2)$ in it signifies the sign structure that appears in the pair correlations, which is crucial for distinguishing DBL from the rest. This is only a concise way to address the differences that arise in different graphs, $e.g.$ Fig.~\ref{fig:DMSP}(b) and Fig.~\ref{fig:DBL}(b).\\
\begin{table}
\begin{tabular}{ | c || c | c | c | } 
\hline
 & SF & DMSP & DBL \\
\hline
\hline
$max(n(q_x,0))$ & $q_x=0$ & $q_x=0$ & $q_x=\pi n_f$ \\
\hline
$O_{DW_2}$ & $0$ & $\neq 0$ & $0$ \\ 
\hline
$O_{twist}$ & $\neq 0$ & $0$ & $\times$ \\
\hline
$sgn(P^{perp}_2)/sgn(P^{para}_2)$ & $1$ &  $1$ &  $-1$ \\
\hline
\end{tabular}
\caption{In the table above, we summarize the behavior of various correlators in different phases. Here, $sgn(\dots)$ picks the sign of the function in its argument. The quantity $sgn(P^{perp}_2)/sgn(P^{para}_2)$ indicates relative sign structure in the pair correlation, which is the basis for d-wave-ness of the DBL, as mentioned in the main text. We note that in case of the DBL, $O_{twist}$ is not meaningful and shows irregular behavior. }
\label{Tab:Ph}
\end{table}

\section{The extended range density-density interaction} \label{DDI}

\begin{figure}
    \centering
    \includegraphics[width=0.45\textwidth]{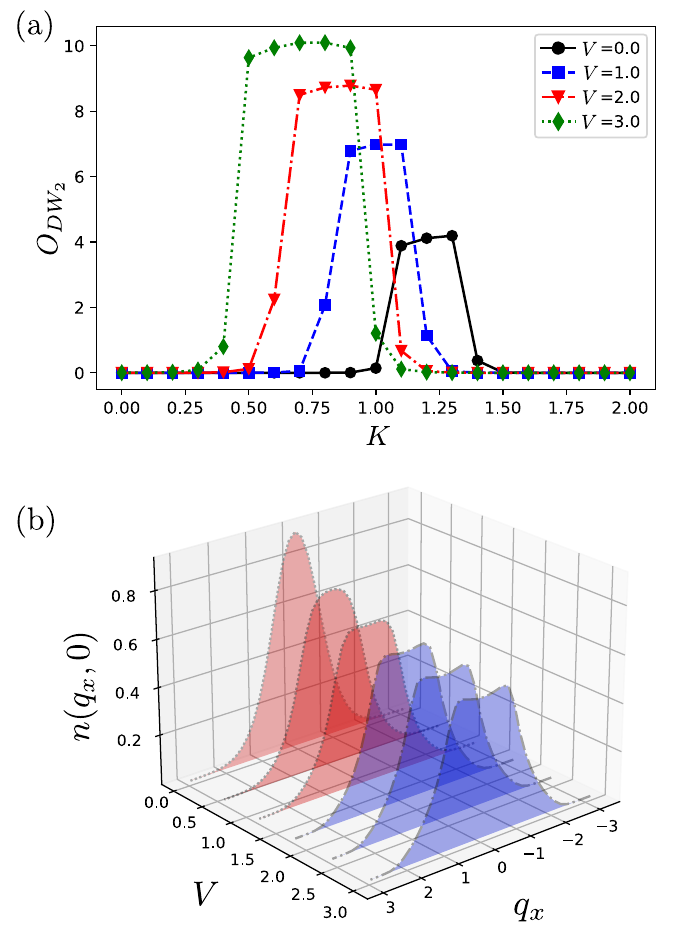}
    \caption{(a) The $O_{DW_2}$ becomes a faithful marker to scrutinize the effect of $V$, as a function of $K$. This quantity is only non-zero in DMSP that appears between SF and DBL. To begin with, for all values of $V$ shown in the plot, the intermediate DMSP sector shifts to a lower $K$, suggesting an early onset of the neighboring DBL phase with increasing $V$. (b) The onset of DBL with increasing $V$ can be explicitly seen, if $n(q_x,0)$ is plotted as a function of $V$. The results are at $K=1.2$. We see the appearance of DBL peaks staring from a DMSP phase at $V=0$ (Fig.~\ref{fig:DMSP}).} 
    \label{fig:ODW2}
\end{figure} 
By now, we have gained a deeper insight regarding the phases present at the $\{ V, U_p  \} \to 0$ limit mimicking a $t-K$ model. Following the motivation laid down in the preceding sections, we make progress towards the finite $V$ limit of the model under study. We initiate by setting $U_p=V$ in Eq.~\eqref{MasHam} to investigate further. This corresponds to a realistic scenario for a two-band system, where NN lattice sites in horizontal and vertical directions are at equal distances. 

One of the most significant results of this paper is summed up in Fig.~\ref{fig:ODW2}. Here, we show how the phase diagram is affected by the presence of density-density interaction of strength $V$. As can be seen in Fig.~\ref{fig:ODW2}(a), when $V$ is increased, the intermediate region shifts to smaller values of $K$. It can be separately checked that the neighboring phases remain the same ($i.e.$ SF and DBL) at finite $V$, within the limits shown in the plot. It implies that finite extended NN and NNN interactions facilitate DBL formation. One gets a deeper insight into this transition by looking into the $n(q_x,0)$ explicitly as a function of increasing $V$. In Fig.~\ref{fig:ODW2}(b), we see the DMSP phase is getting destroyed, and the DBL signatures are becoming prominent. We also note that the peaks are not as prominent as reported in Fig.~\ref{fig:DBL}, which is for $V=0$ and $L_x=48$. However, one can specifically check the pair correlation, other than inspecting the $n(q_x,0)$ closely, to ensure the DBL nature. Other than these, increasing system size also results in sharper peaks. We skip these details to avoid repetitiveness.  \\
\begin{figure} 
    \centering
    \includegraphics[width=0.48\textwidth]{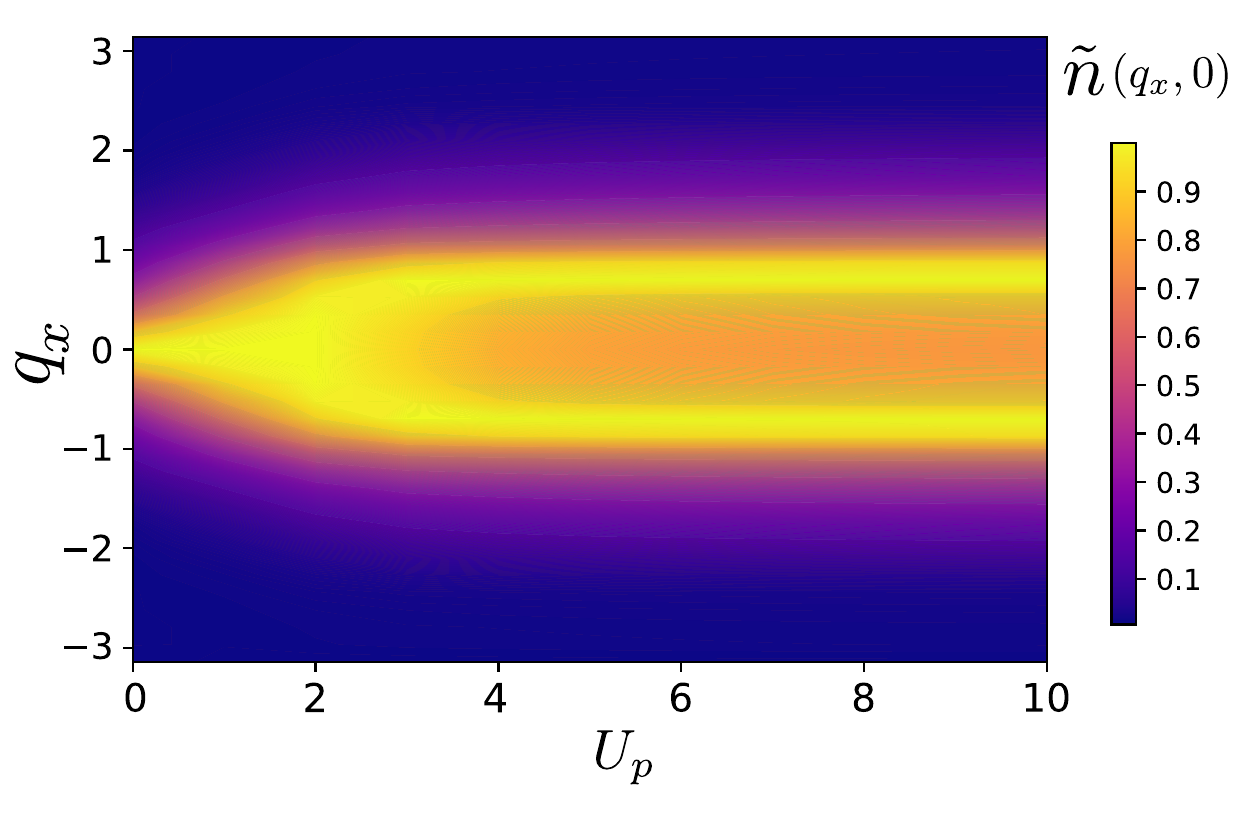}
    \caption{At $V=1.0, K=1.0$ and with $ L_x=36$, we show $\tilde{n}$ (Eq.~\eqref{NqP}) as a function of $U_p$. One can see how the zero-momentum peak gets split with increasing $U_p$. Earlier, in Fig.~\ref{fig:ODW2} we showed how increasing $V$ facilitates the DBL phase. Here we show that the rung interaction $U_p$ can also be tuned to achieve the DBL. } 
    \label{fig:NqUp}
\end{figure}\\
We extend our study further by introducing an asymmetry in NN interaction. The diagonal (NNN) interaction was already unequal to other terms. Now, we create an asymmetry between horizontal and verticle NN interactions. We use $\tilde{n}$ (see Eq.~\eqref{NqP}) to study the emergence of the DBL phase. This function captures the number and positions of peaks in $n(q_x,0)$. Our results show that this quantity accurately tracks the development of double peaks starting from a single peak. The results are shown in Fig.~\ref{fig:NqUp}. In this figure, we notice how dialing up vertical NN interaction stabilizes the DBL further. %This prompts us to explore beyond the two-band model and look into other possible avenues to achieve DBL, as will be shown in the next section.
\section{The rung blockade} \label{StbDBL}
\begin{figure} 
    \centering
    \includegraphics[width=0.48\textwidth]{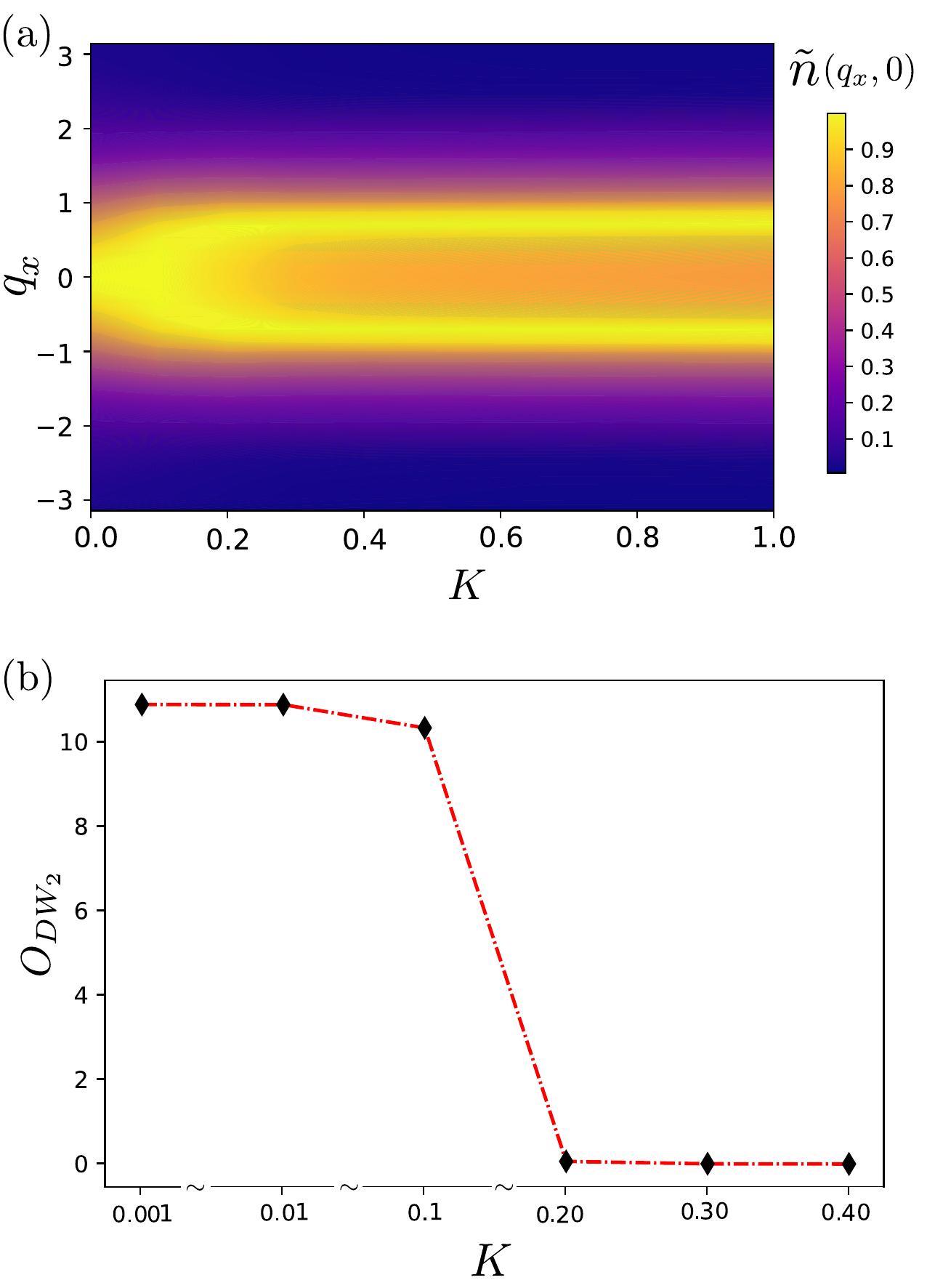}
    \caption{(a) Results for the rung blockade regime. We set V=1.0 and vary $K$ between $0-1$, to understand the scale of $K$, required to achieve a DBL. Here also, like Fig.~\ref{fig:NqUp}, we observe the zero-momenta peak gets split into two. The $K \to 0$ phase is a density wave, as becomes clear from (b), where we show the $O_{DW_2}$: the order parameter for the density wave. The modulation of the density pattern is shown in the Fig.~\ref{fig:DWRB}.} 
    \label{fig:NqRB}
\end{figure}

In the previous section, we have mainly studied the density-density interaction, which is relevant in the context of dipolar particles. Strikingly, we found that these interactions favor DBL formation. In this section, we will focus on another experimentally relevant aspect: the ladder structure produced through synthetic dimension, e.g., through two internal states, with an optical coupling mimicking the transverse hopping. In such a scenario, a single physical site represents the entire rung, and the interaction $U_p$ is actually a local density-density interaction. Therefore, $U_p$ may be much larger than $V$ and $V_d$, up to the limit of a ``rung blockade" given by $U_p \rightarrow \infty$. This prevents double occupation of the rung, a feature that is also present in the proposal of Ref.~\cite{biswas2024}, representing the ladder through a mapping onto three-level ions.

\begin{figure} 
    \centering
    \includegraphics[width=0.44\textwidth]{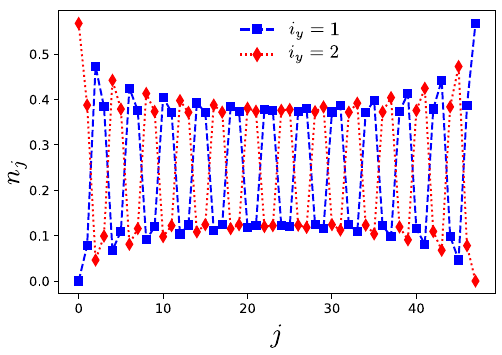}
    \caption{ In the rung blockade regime, a density wave is observed with two-site modulation at low $K$. The above plot is for $L_x=48$ and $K=0.001$. }  
    \label{fig:DWRB}
\end{figure}

\begin{figure} 
    \centering
    \includegraphics[width=0.45\textwidth]{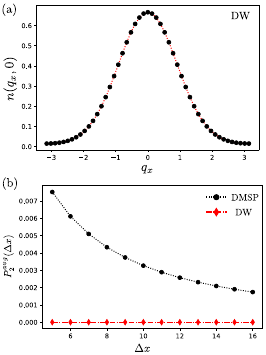}
    \caption{(a) The momentum distribution is peaked at $q_x=0$ and cannot be used to discern DW from DMSP. (b) The quantity $P_2^{avg}$ can distinguish between DW and DMSP, as no pairing mechanism is expected to be present in DW. The plots are for $L_x=48$, and we have used $K=0.001$ and $1.2$ for DW (with blockade) and DMSP, respectively. }  
    \label{fig:DWComp}
\end{figure}

In order to study this limit, we set $U_p/t \sim10^5$ for the purpose of numerical analysis. In this limit, we vary $K$ at fixed $V=1$. This gives us an idea about the scale of $K$ with respect to $V$ when the DBL sets in. The results are depicted in Fig.~\ref{fig:NqRB}. We see that the emergence of DBL is revealed by the appearance of peaks at $\pm \pi/4$ and accompanied by $O_{DW_2}$ going to 0 from a finite value. This is a transition from the density wave (DW) to the DBL phase, as suggested by the $O_{DW_2}$ in Fig.~\ref{fig:NqRB}(b). The density wave pattern, at very low $K$, is displayed in Fig.~\ref{fig:DWRB}. It shows a modulation of single-particle density over two sites as found in Sec.~\ref{PhDMSP} for the DMSP, however, no such pair correlation is present in this case. We use the pairing property to draw a distinction between these two phases. In Fig.~\ref{fig:DWComp}(a), we report the momenta distribution in the DW phase and show that it cannot be used for the purpose of differentiating due to similar qualitative features, whereas in Fig.~\ref{fig:DWComp}(b) we show that $P_2^{avg}$ globally vanishes in DW due to the absence of any two-particle pairing. As a result, $P_2^{avg}$ becomes an ideal candidate for identifying DW separately from DMSP.

We note that both Figs.~\ref{fig:NqUp},~\ref{fig:NqRB} demonstrate the onset of the DBL phase in terms of the appearance of singularities at $\pm \pi/4$. We highlight the fact that, in Fig.~\ref{fig:NqRB}, the value of $K$ is a few fold smaller than that of $V$ for DBL to set in. One can also check that $P_2$ follows expected behavior in DBL. The density wave (DW) phase at $K \to 0$ in Fig.~\ref{fig:NqRB} has to do with the fact that with very strong rung interaction, the blockade term dominates over all other interactions. The presence of DW should be understood on the same footing as that of the extended Bose-Hubbard model in the limit of very high repulsive interaction \cite{Rossini_2012}. One should also recognize that $V_d$ can take a value almost equal to the horizontal NN interaction $V$ if two legs are physically very close to each other. We emphasize that qualitatively, the same result is obtained even if $U_p$ is set to a very high value while maintaining $V_d=V$.

\section{Discussion} \label{Discsn}

We establish the importance of extended range density-density interactions in studying the frustrated Bose ladder. Such processes should be taken into consideration from the point of view of experimental realizations of ring exchange mechanisms, including real or synthetic ladders with dipolar bosons or trapped ions. Strikingly, the presence of density-density interactions is found to enhance the possibility of DBL. In addition to this, the observations on the intermediate DMSP phase (Fig.~\ref{fig:DMSP}) promote a new viewpoint for frustrated ladder systems.

Our principal findings can be categorized into three parts. Firstly, we have identified various phases as a function of ring exchange in the presence of density-density interaction (Fig.~\ref{fig:phase},\ref{fig:ODW2}). In the course of doing so, we noticed the emergence of the novel DMSP phase as a result of ring exchange (Sec.~ \ref{PhDMSP}). Secondly, we have shown that the extended interaction promotes an early commencement of the DBL phase on the $K$ axis (Figs.~\ref{fig:ODW2},\ref{fig:NqUp}). Thirdly, we have reported that the rung blockade (Figs.~\ref{fig:NqRB},\ref{fig:DWRB}) can be utilized to achieve the DBL phase at an appreciably lower value of $K$, compared to the $\{V, U_p \} \to 0$ limit. An important conclusion to be drawn from our findings is that synthetic ladders are particularly promising platforms for the yet outstanding experimental study of DBL physics.

\acknowledgments{ 
We acknowledge the financial support received from the IKUR Strategy under the collaboration agreement between the Ikerbasque Foundation and DIPC on behalf of the Department of Education of the Basque Government. T.G. acknowledges funding by the Department of Education of the Basque Government through the project PIBA\_2023\_1\_0021 (TENINT), and by the Agencia Estatal de Investigación (AEI) through Proyectos de Generación de Conocimiento PID2022-142308NA-I00 (EXQUSMI). This work has been produced with the support of a 2023 Leonardo Grant for Researchers in Physics, BBVA Foundation. The BBVA Foundation is not responsible for the opinions, comments, and contents included in the project and/or the results derived therefrom, which are the total and absolute responsibility of the authors. E.R. acknowledges support from the BasQ strategy of the Department of Science, Universities, and Innovation of the Basque Government. E.R. is supported by the grant PID2021-126273NB-I00 funded by MCIN/AEI/ 10.13039/501100011033 and by ``ERDF A way of making Europe" and the Basque Government through Grant No. IT1470-22. This work was supported by the EU via QuantERA project T-NiSQ grant PCI2022-132984 funded by MCIN/AEI/10.13039/501100011033 and by the European Union ``NextGenerationEU''/PRTR. This work has been financially supported by the Ministry of Economic Affairs and Digital Transformation of the Spanish Government through the QUANTUM ENIA project called – Quantum Spain project, and by the European Union through the Recovery, Transformation, and Resilience Plan – NextGenerationEU within the framework of the Digital Spain 2026 Agenda.}

\bibliography{bib}
\end{document}